\documentclass[5p,times, twocolumn]{elsarticle}

\usepackage{siunitx}
\usepackage{makecell}
\usepackage{hyperref}
\usepackage[ruled,vlined]{algorithm2e}
\usepackage{ragged2e, microtype}
\usepackage{subcaption}
\usepackage{pifont}
\usepackage{amssymb}
\usepackage{float}
\usepackage{array}
\usepackage[table]{xcolor}

\usepackage{etoolbox}
\usepackage{soul} 
\usepackage{graphicx}
\usepackage{multicol}
\usepackage{multirow}

\definecolor{lightgray}{rgb}{0.8, 0.8, 0.8}

\newtoggle{finalPaper}

\setstcolor{red}
\toggletrue{finalPaper} 

\iftoggle{finalPaper} {
	\newcommand{\addtxt}[1]{#1}
	
	\newcommand{\rmvtxt}[1]{}
 	\newcommand{\addtable}[0]{\color{black}}
	}{
	\newcommand{\addtxt}[1]{\textcolor{red}{#1}}
	
	\newcommand{\rmvtxt}[1]{\st{#1}}
 	\newcommand{\addtable}[0]{\color{red}}
}
\begin{document}

\newcommand{\yestick}{{\color{olive}\ding{51}}}
\newcommand{\notick}{{\color{red}\ding{55}}}

\title{Single-board Device Individual Authentication based on Hardware Performance and Autoencoder Transformer Models}

\author[1]{Pedro Miguel {S\'anchez S\'anchez}\corref{cor1}}\ead{pedromiguel.sanchez@um.es}

\author[2]{Alberto {Huertas Celdr\'an}}\ead{huertas@ifi.uzh.ch}

\author[3]{G\'er\^ome Bovet}\ead{gerome.bovet@armasuisse.ch}

\author[1]{Gregorio {Mart\'inez P\'erez}}\ead{gregorio@um.es}

\address[1]{Department of Information and Communications Engineering, University of Murcia, Murcia 30100, Spain}

\address[2]{Communication Systems Group (CSG), Department of Informatics (IfI), University of Zurich UZH, 8050 Zürich, Switzerland}

\address[3]{Cyber-Defence Campus, armasuisse Science \& Technology, 3602 Thun, Switzerland}

\cortext[cor1]{Corresponding author.}

\begin{abstract}

The proliferation of the Internet of Things (IoT) has led to the emergence of crowdsensing applications, where a multitude of interconnected devices collaboratively collect and analyze data. Ensuring the authenticity and integrity of the data collected by these devices is crucial for reliable decision-making and maintaining trust in the system. Traditional authentication methods are often vulnerable to attacks or can be easily duplicated, posing challenges to securing crowdsensing applications. Besides, current solutions leveraging device behavior are mostly focused on device identification, which is a simpler task than authentication. To address these issues, an individual IoT device authentication framework based on hardware behavior fingerprinting and Transformer autoencoders is proposed in this work. \addtxt{To support the design, a threat model details the security problems faced when performing hardware-based authentication in IoT.} This solution leverages the inherent imperfections and variations in IoT device hardware to differentiate between devices with identical specifications. By monitoring and analyzing the behavior of key hardware components, such as the CPU, GPU, RAM, and Storage on devices, unique fingerprints for each device are created. The performance samples are considered as time series data and used to train outlier detection transformer models, one per device and aiming to model its normal data distribution. Then, the framework is validated within a spectrum crowdsensing system leveraging Raspberry Pi devices. After a pool of experiments, the model from each device is able to individually authenticate it between the 45 devices employed for validation. An average True Positive Rate (TPR) of 0.74$\pm$0.13 and an average maximum False Positive Rate (FPR) of 0.06$\pm$0.09 demonstrate the effectiveness of this approach in enhancing authentication, security, and trust in crowdsensing applications.

\end{abstract}


\maketitle

\section{INTRODUCTION}
\label{sec:intro}


The widespread adoption of the Internet of Things (IoT) has led to the emergence of crowdsensing applications, where many IoT devices collaboratively gather and analyze data from the environment \cite{rajendran2018electrosense}. Many of these applications rely on single-board computers due to their reduced price and relatively good performance. These applications offer tremendous potential in diverse domains, such as environmental monitoring, urban planning, healthcare, and transportation. However, ensuring the authenticity and integrity of the data collected by these devices is critical for reliable decision-making and maintaining trust in the system \cite{capponi2019survey}.

The openness and distributed nature of crowdsensing systems make them susceptible to Sybil attacks and collusion among malicious entities \cite{james2020sybil}. Sybil attacks involve adversaries creating multiple fake identities to gain control over the system or manipulate the collected data. Collusion among malicious entities can also lead to coordinated attacks or data manipulation. Implementing identity verification mechanisms, reputation systems, and distributed consensus algorithms is required in order to prevent and detect such attacks \cite{zhong2019connecting}.

Traditional authentication methods for IoT devices, such as cryptographic protocols or unique identifiers, are often susceptible to various attacks and vulnerabilities \cite{wang2020understanding}. Moreover, devices with identical specifications can be easily duplicated or impersonated, posing a significant challenge to maintaining trust and security in crowdsensing applications. To address these limitations, novel approaches are required that leverage the unique characteristics of IoT devices to establish their authenticity. \addtxt{These methods can be seen as an additional layer in the authentication security of IoT scenarios.}

One of the directions proposed in the literature to solve these issues is leveraging hardware manufacturing imperfections in order to uniquely identify each device in the environment \cite{sanchez2020survey}. What elevates the efficiency of this approach is the integration of Machine Learning (ML) and Deep Learning (DL) techniques for the processing of collected hardware behavior data. These cutting-edge computational methodologies facilitate the analysis, classification, and prediction of the enormous amounts of complex, high-dimensional data generated by IoT devices \cite{al2020survey}. Particularly, they can adeptly capture patterns and dependencies in this data, enabling effective anomaly detection and thereby facilitating the identification of devices or activities that deviate from established norms. The combination of hardware manufacturing imperfections and ML/DL techniques has been evidenced to provide remarkable results in the context of device identification \cite{Sanchez2018clock, Sanchez2023methodology}. However, authentication poses a more complex issue: discerning whether a device is authentic or not, but without taking into account the data distributions of other devices. 

Therefore, there are still many challenges present related to hardware-based individual authentication leveraging ML/DL techniques: (\textit{i}) most of the solutions available in the literature cover device identification and not in authentication \cite{sanchez2022specforce}, trying to differentiate a device between a set of known devices instead of uniquely verify its identity; (\textit{ii}) novel DL methods such as attention Transformers have not been applied yet in this field \cite{sanchez2022adversarial}, but could improve current results as it is happening in other fields; (\textit{iii}) solutions are usually implemented in simulated or isolated environments, and not integrated into real-world applications \cite{zhang2019physical}; (\textit{iv}) most of the solutions relying on ML/DL follow a classification-based approach as they focus on identification, which is not practical in dynamic scenarios or when the number of devices is high \cite{al2022cab}.

To solve the previous challenges, the main contributions of the present work are:
\begin{itemize}
    \item A framework that leverages Transformer-based autoencoder models and hardware performance fingerprinting for the individual authentication of single-board computer devices. This framework leverages CPU, GPU, RAM and Storage components to measure their performance and find manufacturing variations that enable the differentiation between devices based on their performance. In this sense, the data from the legitimate device are taken as normal samples modeling its performance distribution, while samples from other devices should be detected as outliers or anomalies.
    
    \item The deployment of the framework in a real-world spectrum crowdsensing platform based on Raspberry Pi devices, namely ElectroSense. In total, 45 devices are utilized in the scenario: 15 Raspberry Pi 4, 10 Raspberry Pi 3, 10 Raspberry Pi 1, and 10 Raspberry Pi Zero. This deployment demonstrates the practical applicability of the framework and its compatibility with different versions of Raspberry Pi devices. It also provides valuable insights into the real-world challenges and considerations in implementing such a sophisticated authentication system, contributing to the broader field of IoT security.

    \item The validation of the framework authentication performance in the deployed scenario. After data collection, an average True Positive Rate (TPR) of 0.74$\pm$0.13 and an average maximum False Positive Rate (FPR) of 0.06$\pm$0.09 are achieved, improving other state-of-the-art models such as LSTM and 1D-CNN networks. This validation not only confirms the effectiveness of the proposed framework but also sets a new benchmark in the field. \addtxt{Besides, a second validation approach details how the solution can be adapted to new device models with different hardware components.} The detailed analysis and comparison with other models provide a comprehensive understanding of the strengths and potential areas for further optimization, paving the way for future research and development in hardware-based authentication. \addtxt{The validation code for the performed experiments is available at \cite{code}.}
\end{itemize}

The remainder of this article is structured as follows. Section \ref{sec:related} gives an overview of hardware-based individual authentication and background on transformer usage for anomaly detection. \addtxt{Section \ref{sec:threat_model} explains the threat model faced by the proposed solution.} Section \ref{sec:authentication} describes the Transformer and hardware-based device fingerprinting solution for individual authentication of single-board devices. Section \ref{sec:validation} gives an overview of the crowdsensing platform employed for validation, the data collection process, and the experimental results when performing the authentication. Finally, Section \ref{sec:conclusions} gives an overview of the conclusions extracted from the present work and future research directions.

\section{RELATED WORK}
\label{sec:related}

\begin{table*}[htpb!]
\centering
\scriptsize
\caption{Comparison of the closest works on ML/DL-focused hardware-based device identification and authentication}
\label{tab:related}
\begin{tabular}{>{\centering\arraybackslash}m{1.6cm}>{\centering\arraybackslash}m{2.5cm}>{\centering\arraybackslash}m{2cm}>{\centering\arraybackslash}m{2cm}>{\centering\arraybackslash}m{1.5cm}>{\raggedright\arraybackslash}m{6cm}}

\textbf{Work} & \makecell[c]{\textbf{Scenario}}& \makecell[c]{\textbf{Approach}} & \textbf{\makecell[c]{Algorthm/Model}}& \textbf{\makecell[c]{N Devices}} & \textbf{\makecell[c]{Results}} \\ \hline  
\hline
\cite{salo2007multi} (2007) & Computer identification & Statistical correlation & Pair-based identification & 38 & Computer identification based on the comparison of three physical oscillators using t-test statistic \\
\hline
\cite{Sanchez2018clock} (2018) & Computer identification & Statistical correlation & Mode-based statistics & 265 & All computers uniquely identified. No effect from CPU load and temperature \\
\hline
\cite{laor2022drawnapart} (2022) & Computer and mobile identification & Classification & CNN & 9 & 95.8\% and 32.7\% accuracy in nine sets of identical devices. Accuracy drops with device rebooting \\
\hline
\cite{Sanchez2023methodology} (2023) & IoT device identification & Classification & XGBoost & 25 & 91.92\% average TPR. No effects from temperature changes and device rebooting \\
\hline
\cite{sanchez2022adversarial} (2023) & IoT device identification & Classification & LSTM + 1D-CNN & 45 & 0.96 average F1-Score. Resilience to temperature and ML/DL evasion attacks.\\
\hline
\textbf{This work (2023)} & IoT device authentication & Anomaly Detection & Transformer & 45 & All devices authenticated. 0.74 average TPR and 0.06 average maximum FPR \\
\hline
\end{tabular}%
\end{table*}

This section reviews the key literature relevant on individual device authentication through hardware performance fingerprinting and transformer-based anomaly detection.

\subsection{Individual device authentication and identification}

The present work focuses on hardware-based single-board device authentication using the performance behavior of the components self-contained in the device and anomaly detection DL algorithms. Arafin and Qu \cite{arafin2021hardware} discussed several examples of hardware-based authentication that use memory access latency, instruction execution latency, and clock skew to authenticate devices, users, and broadcast signals used for navigation. In \cite{Sanchez2023methodology}, the authors compared the deviation between the CPU and GPU cycle counters in Raspberry Pi devices to perform individual identification of 25 devices. The identification was performed using XGBoost, achieving a 91.92\% True Positive Rate (TPR). In continuing work \cite{sanchez2022adversarial}, the same authors improved the results to an average F1-Score of +0.96 and a minimum TPR of 0.8 using a time series classification approach based on LSTM and 1D-CNN combination. Similarly, \cite{laor2022drawnapart} performed identical device identification using GPU performance behavior and ML/DL classification algorithms. Accuracy between 95.8\% and 32.7\% was achieved in nine sets of identical devices, including computers and mobile devices. Sanchez-Rola et al. \cite{Sanchez2018clock} identified +260 identical computers by measuring the differences in code execution performance. They employed the Real-Time Clock (RTC), which includes its own physical oscillator, to find slight variations in the performance of each CPU. In \cite{salo2007multi}, the author compared the drift between the CPU time counter, the RTC chip, and the sound card Digital Signal Processor (DSP) to identify identical computers. Other works have also explored hardware-based authentication applications using physical properties of computing hardware such as main memory, computing units, and clocks. Shrivastava et al. \cite{shrivastava2022high} proposed a high-performance Field Programmable Gate Arrays (FPGA) based secured hardware model for IoT devices using the Advanced Encryption Standard (AES) algorithm. They compared the performance of two FPGAs and found that the Spartan-6 FPGA provides better throughput and less time delay for IoT devices. 

Other works have explored the usage of Physical Unclonable Functions (PUFs) for IoT device identification \cite{shamsoshoara2020survey}. However, PUFs are out of the scope of this work, as it is centered on hardware behavior fingerprinting based on device performance, avoiding the usage of new hardware elements or the modification of the device specifications.

\addtxt{Finally, some solutions are also available in the industry, leveraging hardware characteristics for IoT device identification. Numerous hardware-based authentication solutions for IoT devices have been introduced to enhance security. Intel Enhanced Privacy ID (EPID) provides a mechanism for device authentication while ensuring privacy, making certain that devices connecting to networks are genuine Intel products \cite{intel_epid}. ARM TrustZone technology partitions devices into secure and non-secure zones, offering a foundational layer for security solutions \cite{arm_trustzone}. Cisco Trust Anchor module (TAm) embeds a hardware module in products to guarantee device integrity and authenticity right from manufacturing to deployment \cite{cisco_tam}. Microsoft has ventured into this space with Azure Sphere, which incorporates custom silicon security technology for comprehensive IoT security \cite{azure_sphere}. NXP A71CH is a secure element designed to provide a root of trust at the integrated circuit level for IoT devices \cite{nxp_a71ch}. Infineon OPTIGA Trusted Platform Module (TPM) offers hardware-based security functions, facilitating device authentication \cite{infineon_optiga}. Microchip CryptoAuthentication devices are tailored to protect against various security threats by offering robust cryptographic solutions \cite{microchip_cryptoauth}. Rambus CryptoManager IoT Device Management is a turnkey solution designed to provide end-to-end security, including device attestation and hardware-based security \cite{rambus_cryptomanager}. Lastly, GlobalPlatform Device Trust Architecture (DTA) standardizes the use of secure components in IoT devices to protect digital services and data \cite{globalplatform_dta}. While hardware-based industrial authentication solutions for IoT devices bolster security, they come with challenges. These include higher costs, increased deployment complexity, computational overhead on devices, reduced flexibility for updates, scalability concerns in vast networks, vulnerabilities to physical attacks, supply chain integrity issues, interoperability problems among different manufacturers, potential long-term hardware degradation affecting performance, and the risk of vendor lock-in due to proprietary solutions.}

\tablename~\ref{tab:related} compares the closest works in the literature with the present one. Although several works have worked in the combination of ML/DL techniques and hardware fingerprinting for device identification, a notable gap persists in the literature with respect to addressing the unique challenges of device authentication via an anomaly detection approach. Contemporary studies have primarily employed classification models, which serve to identify devices from a set pool of labels. However, these models are inadequate for the authentication problem. The task of authentication involves more than simple device recognition - it requires a system capable of detecting deviations from an expected hardware behavior, a task for which anomaly detection models, rather than traditional classification models, are better suited. Consequently, there is a significant need to investigate the potential of DL-based anomaly detection models, such as Transformer models, in the realm of device authentication.

\subsection{Transformer-based anomaly detection in IoT security}

The application of Transformer models in anomaly detection has recently gained momentum, recognizing their ability to extract meaningful features from sequential data effectively. Anomaly detection in time-series data, in particular, has seen significant advancements through the adoption of Transformer models \cite{choi2021deep}. Their proficiency in capturing temporal dynamics makes them an excellent choice for tasks that involve detecting irregularities in time-bound sequences \cite{tuli2022tranad}.

In the field of IoT security, Transformer-based autoencoders have been employed to address high-dimensional and complex dependencies issues by leveraging the self-attention mechanism and the encoder-decoder architecture. Chen et al. \cite{chen2021learning} proposed a framework called GTA that learns a graph structure among sensors and applies graph convolution and Transformer-based modeling to detect anomalies in multivariate time series. Kozik et al. \cite{kozik2021new} proposed a hybrid time window embedding method with a Transformer-based classifier to identify compromised devices in IoT-networked environment. Tuli et al. \cite{tuli2022tranad} proposed TranAD, a deep Transformer network that uses attention-based sequence encoders to perform anomaly detection and diagnosis for IoT data streams. These works demonstrate the effectiveness and efficiency of Transformer-based models for anomaly detection in IoT security.

However, the performance of Transformer-based anomaly detection in individual device authentication has not been explored yet, remaining as a practical field where the performance of these novel models can improve the state-of-the-art approaches.

\section{\addtxt{THREAT MODEL}}
\label{sec:threat_model}

\addtxt{The primary concern in the single-board device authentication scenario is an adversarial actor attempting to integrate an unauthorized device into a sensitive setting, like an industry, by masquerading as or impersonating a legitimate device. This threat can be approached from multiple angles:}

\begin{itemize}
\item \addtxt{\textit{TH1. Device impersonation} \cite{marabissi2022iot}. The foremost security challenge is when an adversarial entity substitutes a genuine device with a malicious device that mirrors its software characteristics. In this case, the adversary deploys identical legitimate software credentials but incorporates malevolent processes and features.}

\item \addtxt{\textit{TH2. Sybil} \cite{rajan2017sybil}. A singular device (or multiple) might attempt to create numerous authentications to transmit deceptive data from many mimicked devices. The vulnerability of a system to Sybil attacks hinges on (i) the simplicity of creating authentications; (ii) whether the system uniformly handles all entities, and (iii) the extent to which the system approves of entities lacking a trust linkage to a recognized trustworthy entity.}

\item \addtxt{\textit{TH3. Replay Attacks} \cite{feng2017replay}. Attackers can capture authentication tokens or messages and replay them later to gain unauthorized access or to disrupt the network operations.}

\item \addtxt{\textit{TH4. Physical Attacks} \cite{stellios2021assessing}. Many IoT devices are deployed in environments that lack stringent security, leaving them vulnerable to physical tampering. Malicious actors can directly access these devices to extract sensitive cryptographic keys or implant malicious hardware/software components. The direct physical access grants attackers a high level of control, making these attacks particularly devastating.}

\item \addtxt{\textit{TH5. Advanced persistent threat} \cite{chen2022machine}. This threat emerges as an outcome of the preceding one. A rogue device set up in the environment may be capable of extracting data from the situation and other devices or initiating more aggressive assaults like vulnerability scanning and/or Denial of Service (DoS) attacks. Additionally, contemporary attacks typically incorporate evasion methods that conceal their operations from software-centric behavior observation security mechanisms \cite{li2020adversarial}.}
\end{itemize}

\addtxt{Traditional software-based authentication methods, while effective in some scenarios, have shown vulnerabilities in the face of sophisticated threat models where certificates or software identifiers can be cloned. Therefore, solving the previous threat model is the main objective of the proposed solution, complementing traditional authentication systems based on software. By capitalizing on inherent cycle skew and performance disparities in hardware -even among identical IoT devices- this approach can establish a unique, tamper-resistant identity for each device. These intrinsic hardware traits offer not just a shield against software-based incursions but also a robust defense against physical intrusions. Additionally, by folding hardware performance metrics into the authentication matrix, the solution can seamlessly cater to the diverse performance spectra of IoT devices, facilitating efficient authentication processes, even for those with resource constraints.}

\addtxt{In order to solve the threats identified in this work, it is assumed that even if the device is malicious, the control over it is maintained by its legitimate administrator and the authentication tasks can be executed. This condition guarantees that device management is maintained during a possible attack. If this control is lost, it would be assumed that the device is infected or has some error.}

\section{INDIVIDUAL DEVICE AUTHENTICATION FRAMEWORK}
\label{sec:authentication}

This section elucidates the DL framework implemented for the purpose of hardware performance fingerprinting. The framework performs device fingerprinting based on performance deviations that show hardware manufacturing imperfections. An autoencoder Transformer model, a state-of-the-art approach in DL-based time series processing, is leveraged for the authentication of individual devices. 

\begin{figure*}[htpb!]
    \centering \includegraphics[width=\textwidth]{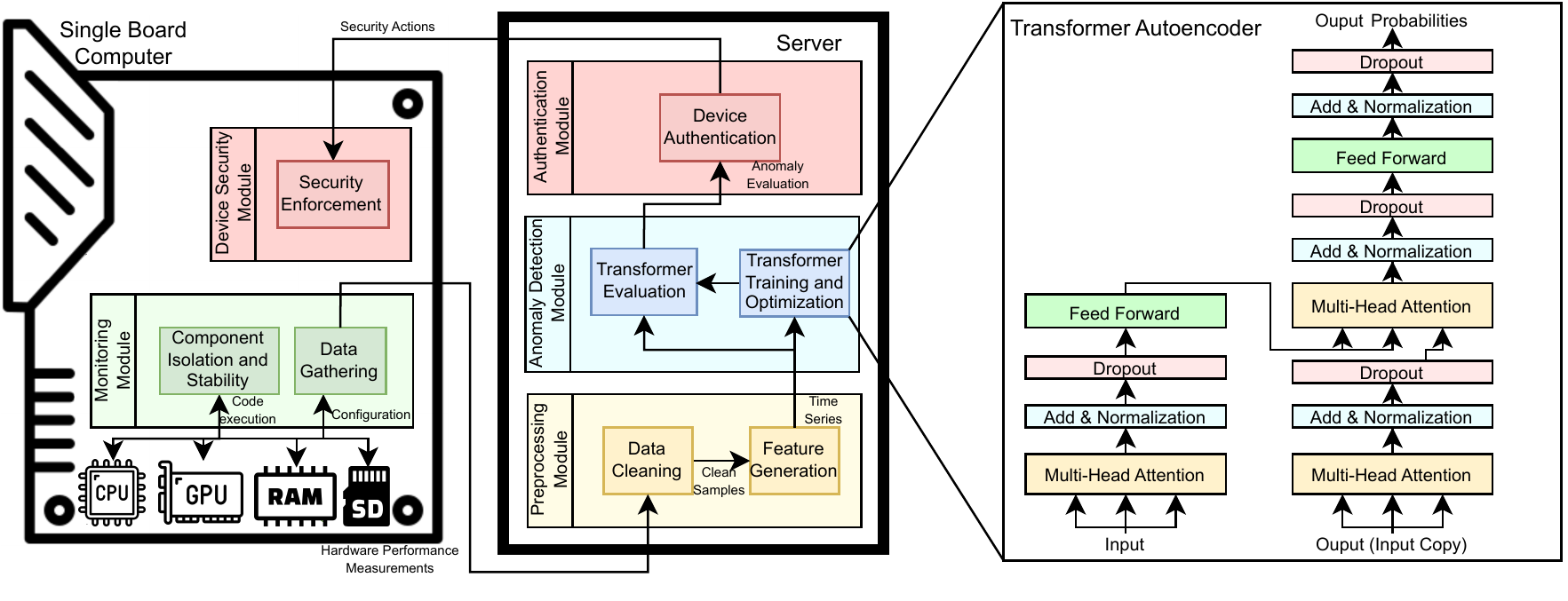}
    \caption{Individual device authentication framework.}
    \label{fig:framework}
\end{figure*}

The framework is designed in a modular manner, where different components are combined in a stacked layout, from the hardware behavior monitoring to the DL-based evaluation and authentication. Due to the reduced processing capabilities of single-board computers, the framework follows a client-server architecture, where the components related to data collection and device configuration are deployed locally in the device, and the server processes the data and performs the model training and evaluation. \figurename~\ref{fig:framework} illustrates the different modules composing the framework and the pipeline followed by the data until an authentication decision is made. Five modules compose the framework: \textit{(i)} Monitoring, \textit{(ii)} Preprocessing, \textit{(iii)} Anomaly Detection, \textit{(iv)} Authentication, and \textit{(v)} Device Security.

\subsection{Monitoring Module}

The \textit{Monitoring Module} is in charge of the interaction with the hardware components and the monitoring of their performance. Besides, it sends the collected data to the server for its processing and evaluation. It contains two components: \textit{Component Isolation and Stability} and \textit{Data Gathering}.

\subsubsection{Component Isolation and Stability} One of the key conditions to perform fingerprinting based on hardware performance is to ensure that the components selected for monitoring are running under stable conditions that enable the characterization of the small performance variations in the components due to manufacturing imperfections \cite{Sanchez2023methodology}. Therefore, this component is in charge of configuring the CPU, GPU, RAM and SD Card, the selected hardware components. It sets fixed running frequency for the components, isolate the components to avoid kernel interruptions, and disables some component optimizations that might affect the stability of the performance, such as memory address randomization.

\subsubsection{Data Gathering} This component is in charge of collecting the performance measurements by executing different tasks in the selected hardware components. In the case of single-board computers, the available hardware elements are the CPU, GPU, RAM and storage (typically SD card). As proposed in the literature \cite{Sanchez2023methodology}, the hardware monitoring is done by using the in-device elements as a reference for the performance measurements. For example, GPU performance is measured in CPU cycles, and CPU performance when executing a code is measured using the elapsed GPU cycles. The reasoning for this approach is that the component itself is not able to measure the deviations in its performance specification without an external cycle or time counter.

\subsection{Preprocessing Module}

The \textit{Preprocessing Module} plays the pivotal role of a bridge between the raw data gathered by the \textit{Monitoring Module} and the \textit{Anomaly Detection Module}, where the data is employed to train the DL models and evaluate the device. The main tasks of this module encompass data cleaning and feature generation.

\subsubsection{Data Cleaning} This component is responsible for filtering and cleaning the raw performance metrics. Any missing, inconsistent, or erroneous data are identified and filtered, thus preparing the dataset for further processing.

\subsubsection{Feature Generation} This component focuses on feature extraction and engineering based on the cleaned data. First, it performs normalization of each one of the metrics gathered. Afterward, it is in charge of transforming the raw data into a format suitable for the Transformer model. A key aspect of this process is the concatenation of samples into groups of vectors, which facilitates time series-based analysis.

\subsection{Anomaly Detection Module}

The \textit{Anomaly Detection Module} is the heart of the authentication framework, tasked with training and evaluating the Transformer-based autoencoder model. The Transformer-based autoencoder is a variant of the Transformer model, which was originally proposed for natural language processing tasks. The key component of the Transformer architecture is the self-attention mechanism, which models the interactions between the elements in the input sequence \cite{vaswani2017attention}. More in detail, the self-attention mechanism computes a weighted sum of the input elements for each position in the sequence. The weight assigned to each input element is determined by its relevance to the position being considered. Formally, the self-attention can be computed as follows:

\begin{equation}
\text{Attention}(Q, K, V ) = \text{softmax}\left(\frac{QK^T}{\sqrt{d_k}}\right)V
\end{equation}

where $Q$, $K$, and $V$ are matrices representing the queries, keys, and values, respectively, and $d_k$ is the dimensionality of the keys. In multi-head attention, this operation is done $h$ times with different learned linear projections of the original $Q$, $K$, and $V$ matrices. 

In the autoencoder variant of the Transformer model, the same sequence is provided as both the input and the target output of the model. The Transformer-based autoencoder learns to reconstruct the input sequence, which allows it to capture the underlying structure of the sequence data.

The encoder and decoder are both composed of several identical layers. Each layer contains two sub-layers: a multi-head self-attention mechanism and a position-wise fully connected feed-forward network, using ReLU as activation function. The output of each sub-layer is then passed through a residual connection and layer normalization.

In the context of device authentication, the Transformer-based autoencoder is trained to reconstruct the normal behavior of each device. Once the model is trained, it can be used to detect anomalies by comparing the reconstruction error of a new sequence with a predefined threshold. A high reconstruction error indicates that the new sequence is significantly different from the normal behavior, which could suggest a possible intrusion.

The two components forming this module, in charge of the Transformer-based autoencoder training for each device, are:

\subsubsection{Transformer Training and Optimization}

This component takes the processed data and trains a Transformer model for each device. This model, adept at reconstructing input data, establishes a profile of standard device behavior, thereby becoming proficient at detecting anomalies or deviations from the norm. This phase also involves the optimization of model parameters for each device independently to ensure the best performance. Then, the best model for each device is stored to be later used. The training and optimization process is iterative and may require several exploratory iterations to find the combination that meets all the properties needed in the generated fingerprint.

\subsubsection{Transformer Evaluation}

Upon completion of the training phase, the model is subject to deployment for live data evaluation. The model predictive capability is tested against the values collected from the device after deployment. Then, the output of the Transformer will be employed in the \textit{Authentication Module} to determine if a device is the legitimate one and grant allow him to remain deployed in the network. Any deviations from the established profile may trigger further investigation or immediate action, depending on the \textit{Device Security Module}.

\subsection{Authentication Module}

The \textit{Authentication Module} makes the final decision regarding device authentication based on the evaluation results coming from the previous module. It integrates the anomaly detection results with other contextual information, such as device history or network behavior, to make a more informed decision. This module may also include additional verification steps or multi-factor authentication to enhance security.

\subsubsection{Device Authentication}

This component is charged with the essential task of making the final authentication decision based on the anomaly detection results. Anomalies, interpreted as potential indications of device tampering or misuse, inform the authentication decision. A device may be authenticated and granted network access, or it may be rejected, depending on the analysis of these anomalies.

\subsection{Device Security Module}

The \textit{Device Security Module} serves as an additional layer of security, overseeing the enforcement of security measures. It works in conjunction with the \textit{Authentication Module} to provide a comprehensive security solution for IoT devices after authentication.

\subsubsection{Security Enforcement}

This component ensures the enforcement of necessary security rules or protocols based on the \textit{Authentication Module} decision. If a device is authenticated, it is granted access to the network. If a device is deemed unauthenticated, this component ensures the device is isolated from the network, safeguarding the integrity of the IoT system. This module also reports any security issues, such as repeated authentication failures, to a central authority for further investigation. Moving target defense (MTD) techniques are a suitable approach for this module, as they focus on changing the device configuration according to the mitigation actions required. Some examples of these techniques is the removal of files, dynamic network connection filtering, among others.

\section{FRAMEWORK VALIDATION}
\label{sec:validation}

This section succinctly lays out the overall validation methodology, from leveraging the ElectroSense spectrum crowdsensing platform to data collection and preprocessing crucial for the analysis. The specifics of data gathering and the processes of cleaning, normalization, and transformation are explained. Finally, the Transformer-based Anomaly Detection model approach is validated in this real-world scenario, measuring its effectiveness. Note that the validation focuses on the data collection, monitoring, and DL parts of the framework. The development of advanced authentication rules and security measures is out of the scope of this work.

\subsection{ElectroSense spectrum crowdsensing platform}

The IoT spectrum sensors utilized in this research are a part of the ElectroSense network \cite{rajendran2018electrosense}, an open-source, crowdsensing platform that collects radio frequency spectrum data with the aid of low-cost sensors. The platform, which capitalizes on a collaborative crowdsensing approach, enables the monitoring and collection of spectrum data. The core of this platform is the Raspberry Pi, a compact and cost-effective single-board computer, that when attached to software-defined radio kits and antennas can function as a versatile spectrum sensor. Such assembly of spectrum sensors by individual users contributes to the broad reach and comprehensive data collection capability of the ElectroSense platform.

\begin{figure}[htpb!]
    \centering \includegraphics[width=\columnwidth]{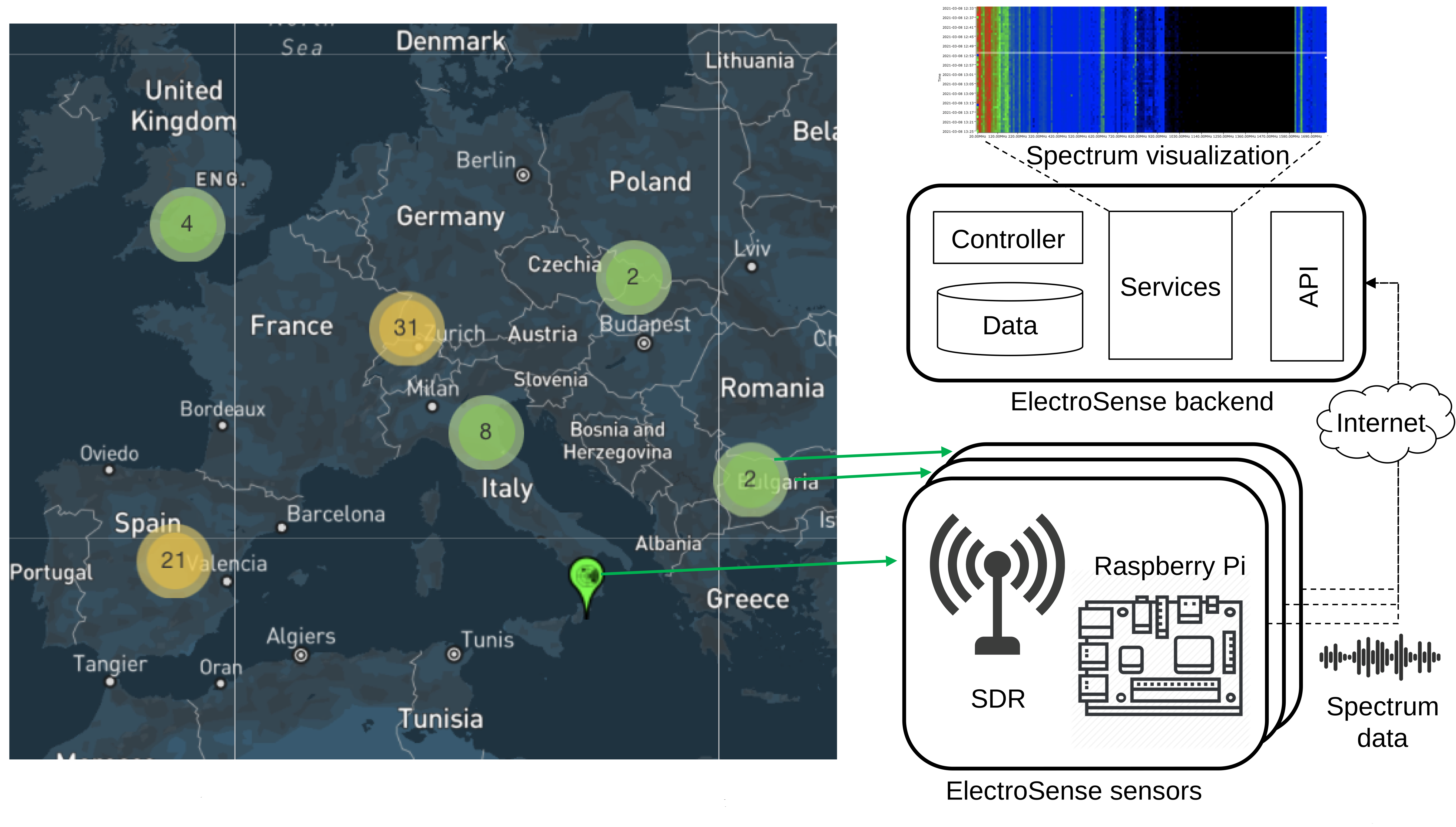}
    \caption{ElectroSense crowdsensing platform diagram.}
    \label{fig:electrosense}
\end{figure}

Once the sensors have collected the data, it is then sent to the ElectroSense backend platform, which is responsible for its storage, processing, and analysis. This meticulous processing and analysis facilitate the provision of a suite of services. These services extend beyond mere spectrum occupancy monitoring, delving into areas such as transmission optimization and decoding. This range of services provided by ElectroSense not only bolsters the understanding of spectrum utilization but also opens up avenues for innovative optimization and enhancement strategies in the field of IoT. \figurename~\ref{fig:electrosense} depicts a diagram of the ElectroSense platform.

For validation, numerous Raspberry Pi devices from different models are deployed in the crowdsensing platform in order to validate the proposed authentication framework. More in detail, the devices deployed are 15 Raspberry Pi 4 Model B, 10 Raspberry Pi 3 Model B+, 10 Raspberry Pi Model +, and 10 Raspberry Pi Zero.

\subsection{Data Gathering and Preprocessing}

The first step in the validation process is to obtain the hardware performance data from each device and preprocess it in order to be fed into the Transformer models.

\subsubsection{Data Gathering} The assembly of individual device authentication premised on hardware behavior hinges on the ability to monitor imperfections inherent in the device chips for subsequent evaluation. As outlined in Section \ref{sec:related}, previous studies have primarily tackled this task by contrasting components featuring different base frequencies or crystal oscillators since deviations in these components performance can be discerned directly from the device.

To construct the framework for individual device authentication, it was necessary to compile a dataset that utilizes metrics pertinent to the hardware components inherent in certain devices. This dataset has been christened LwHbench, and additional details can be found in \cite{sanchez2022lwhbench}. In this context, the dataset gathered performance metrics from the CPU, GPU, Memory, and Storage of 45 Raspberry Pi devices of diverse models over a span of 100 days. Various functions were executed in these components, employing other hardware elements (operating at differing frequencies) to measure performance. \tablename~\ref{tab:features_dataset} provides a summary of the functions that were monitored. These functions embody a set of common operations carried out in every component, aiming to gauge their performance. It is worth mentioning that additional analogous operations could be utilized during the data gathering process. In total, 215 features formed each one of the collected data vectors.

\begin{table}[htpb!]
\centering
\scriptsize
\caption{LwHBench dataset features \cite{sanchez2022lwhbench}.}
\begin{tabular}{ >{\raggedright\arraybackslash}m{1.5cm} >{\raggedright\arraybackslash}m{1.5cm} >{\raggedright\arraybackslash}m{4.55cm} }
\textbf{Component} & \textbf{Function} & \textbf{Feature Under Observation} \\
\hline
\hline
\textbf{-} & timestamp & Unix timestamp \\
& temperature & Core temperature of the device \\
\hline
\textbf{CPU} & 1 s sleep & Elapsed GPU cycles during 1s of CPU sleep \\
& 2 s sleep & Elapsed GPU cycles during 2s of CPU sleep \\
& 5 s sleep & Elapsed GPU cycles during 5s of CPU sleep \\
& 10 s sleep & Elapsed GPU cycles during 10s of CPU sleep \\
& 120 s sleep & Elapsed GPU cycles during 120s of CPU sleep \\
& string hash & Elapsed GPU cycles during computation of a fixed string hash \\
& pseudo random & Elapsed GPU cycles while generating a software pseudo-random number \\
& urandom & Elapsed GPU cycles while generating 100 MB using \textit{/dev/urandom} interface \\
& fib & Elapsed GPU cycles while calculating the 20th Fibonacci number using the CPU \\
\hline
\textbf{GPU} & matrix mul & Time taken by CPU to execute a GPU-based matrix multiplication \\
& matrix sum & Time taken by CPU to execute a GPU-based matrix summation \\
& scopy & Time taken by CPU to execute a GPU-based graph shadow processing \\
\hline
\textbf{Memory} & list creation & Time taken by CPU to generate a list with 1000 elements \\
& mem reserve & Time taken by CPU to fill 100 MB in memory \\
& csv read & Time taken by CPU to read a 500 kB \textit{csv} file \\
\hline
\textbf{Storage} & read x100 & 100 measurements of CPU time for 100 kB storage read operations \\
& write x100 & 100 measurements of CPU time for 100 kB storage write operations \\
\hline
\end{tabular}
\label{tab:features_dataset}
\end{table}

The final dataset contains the following samples per device model: 505584 samples collected from 10 RPi 1B+ devices, 784095 samples from 15 RPi4 devices, 547800 samples from 10 RPi3 devices, and 548647 samples from 10 RPiZero devices. To collect the data, an array of countermeasures were implemented to mitigate the effect of noise introduced by other processes operating in the devices: Component frequency was kept constant, kernel level priority was enforced, the code was executed in an isolated CPU core (in multi-core devices), and memory address randomization was disabled. Moreover, the dataset was compiled under a variety of temperature conditions, facilitating the analysis of the influence this environmental feature has on component performance.

\subsubsection{Preprocessing} 
In the preprocessing stage, the time series were generated by applying a time window over the collected samples, combining them into groups of 10 to 100 vectors. This method of grouping facilitates the implementation of time series Deep Learning (DL) approaches and is adjusted to other literature works \cite{sanchez2022adversarial}. These models possess the ability to uncover intricate trends within the data, potentially leading to superior results compared to the standalone processing and evaluation of individual samples. Moreover, it also permits the utilization of attention models such as Transformers, which currently represent the pinnacle of performance in this field. For data normalization, \textit{QuantileTransformer} \cite{ahsan2021effect} was utilized, given the variable data distributions originating from the differing hardware capabilities of each device model. The division of the data for model training and validation purposes consisted of 70\% and 10\% of the total, leaving the remaining 20\% for testing. In order to minimize the potential impact of vector order correlations on the results, the splitting of training, validation, and test sets was performed without shuffling the samples.

\subsection{Transformer-based Anomaly Detection Validation}


As detailed in Section~\ref{sec:authentication}, the proposed Transformer approach performs hyperparameter tuning personalized for each device. Besides, other state-of-the-art DL architectures for anomaly detection in time series are tested to compare their performance to the Transformer. The tested networks are LSTM, 1D-CNN, and a combination of both of these layouts. \tablename~\ref{tab:clasif_alg_hyp} provides a comprehensive overview of the examined algorithms along with their corresponding hyperparameters. For validation, a server equipped with AMD EPYC 7742 CPU, NVIDIA A100 GPU, and 180 GB of RAM is employed, and the models are implemented using \textit{Keras} library.
 
\begin{table}[htpb!]
	\caption{Anomaly detection time series models and hyperparameters tested.}
	\centering
    \scriptsize
    \begin{tabular}{m{1.15cm}m{6.9cm}}
        \textbf{Model} & \textbf{Hyperparameters}\\
        \hline
        \hline
        \makecell[l]{General\\Parameters} & $epochs=[10,20,50], batch\_size=[32,128,256,512]$\\
        \hline
        1D-CNN & \makecell[l]{$filters=[16,32,64,128],kernel\_size=[3,5,7],$\\$n\_layers=[1,2,3]$}\\
        \hline
        LSTM & \makecell[l]{$neurons=[10,100], n\_layers=[1,2,3],$}\\
        \hline
        \makecell[l]{LSTM\_\\1D-CNN} & \makecell[l]{$input\_layers=[2,3], cnn\_filters=[16,32,64,128],$ \\$cnn\_kernel\_size=[3,5,7], lstm\_neurons=[10,100]$\\$n\_layers=[1,2,3]$}\\
        \hline
        Transformer & \makecell[l]{$dff=[32,64,128,256,1024], num\_layers=[1,2,3]$\\}\\
        \hline
    \end{tabular}
    \label{tab:clasif_alg_hyp}
\end{table}{}

In the case of the LSTM and 1D-CNN models, the time series concatenation only achieved good results when using groups of 10 vectors or smaller due to their limited memory capabilities. In contrast, the Transformer achieved good results with all the sliding window lengths from 5 to 100, with the best results obtained with 100 vectors per sliding window.

To set the anomaly detection threshold in the reconstruction of the samples fed to the autoencoder models, the 10\% of the reconstruction error in the training samples is chosen as the boundary between anomaly and normal sample. Then, the validation set is employed for the hyperparameter selection by choosing the model with the higher TPR.

\subsubsection{Authentication Performance}
For the authentication capabilities evaluation, the strategy followed is one-vs-all, where the trained transformer model evaluates the test set of the source device (normal samples) but also the test sets of the rest of the devices (anomalies or outliers). Then, the True Positive Rate (TPR) of the legitimate device is compared with all the False Positive Rates (FPRs) of the rest of the devices, checking that the TPR value is greater than all the FPRs. Note that for this approach, different data normalizations should be performed in the test sets depending on which device is employed for training as the training data distribution changes.

\tablename~\ref{tab:anom_results} shows the results of the one-vs-all authentication tests. It can be seen how only the Transformer-based approach is able to authenticate all the devices successfully. Although their average TPR is higher, LSTM and 1D-CNN networks only can identify some of the devices, offering a much lower difference between the average TPR and maximum FPR. This occurs because the FPR is much more variable in these models, and many models have a high FPR when evaluating data from other devices, while the FPR variability is smaller in the Transformer models.

\begin{table}[htpb!]
	\caption{Anomaly detection time series models results.}
	\centering
    \scriptsize
    \begin{tabular}{m{1.7cm}m{1.3cm}m{1.5cm}m{1.1cm}m{1.1cm}}
        \makecell[l]{\textbf{Model}} & \textbf{Best window size} &\textbf{Devices Authenticated} & \textbf{Avg. TPR} & \textbf{Avg. Max. FPR} \\
        \hline
        \hline
        1D-CNN & \makecell[c]{10} & \makecell[c]{32} & \makecell[c]{0.88$\pm$0.06} & \makecell[c]{0.67$\pm$0.29}  \\
        \hline
        LSTM & \makecell[c]{10} & \makecell[c]{38} & \makecell[c]{0.85$\pm$0.09} & \makecell[c]{0.53$\pm$0.19}  \\
        \hline
        \makecell[l]{LSTM\_1D-CNN} & \makecell[c]{10} & \makecell[c]{35} & \makecell[c]{0.88$\pm$0.08} & \makecell[c]{0.59$\pm$0.22} \\
        \hline
        \rowcolor{lightgray}
        Transformer & \makecell[c]{100} & \makecell[c]{45} & \makecell[c]{0.74$\pm$0.13} & \makecell[c]{0.06$\pm$0.09} \\
        \hline
    \end{tabular}
    \label{tab:anom_results}
\end{table}{}

\begin{figure}[htpb!]
    \centering \includegraphics[width=\columnwidth]{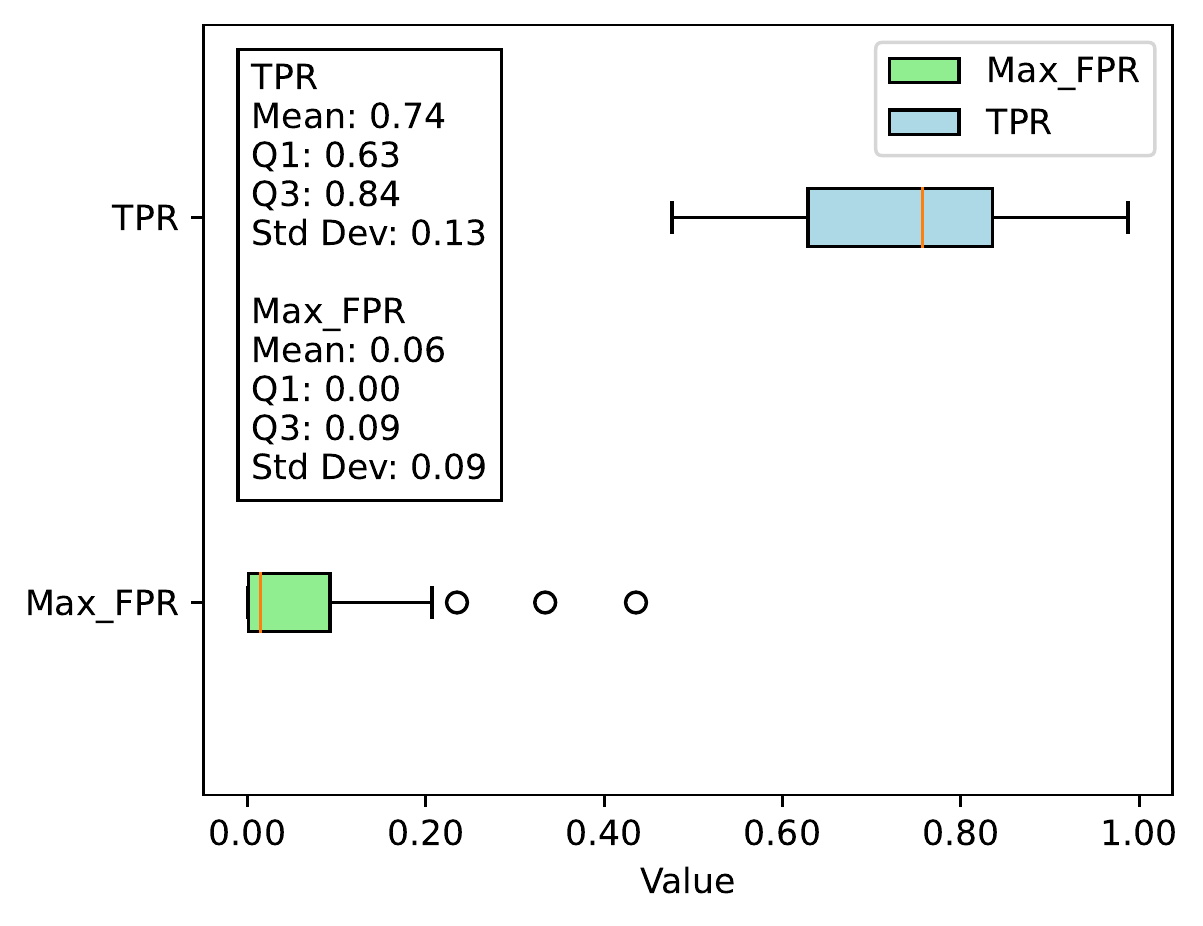}
    \caption{TPR and maximum FPR distributions of the Transformer autoencoder.}
    \label{fig:distributions}
\end{figure}

\begin{figure*}[htpb!]
    \centering \includegraphics[width=\textwidth]{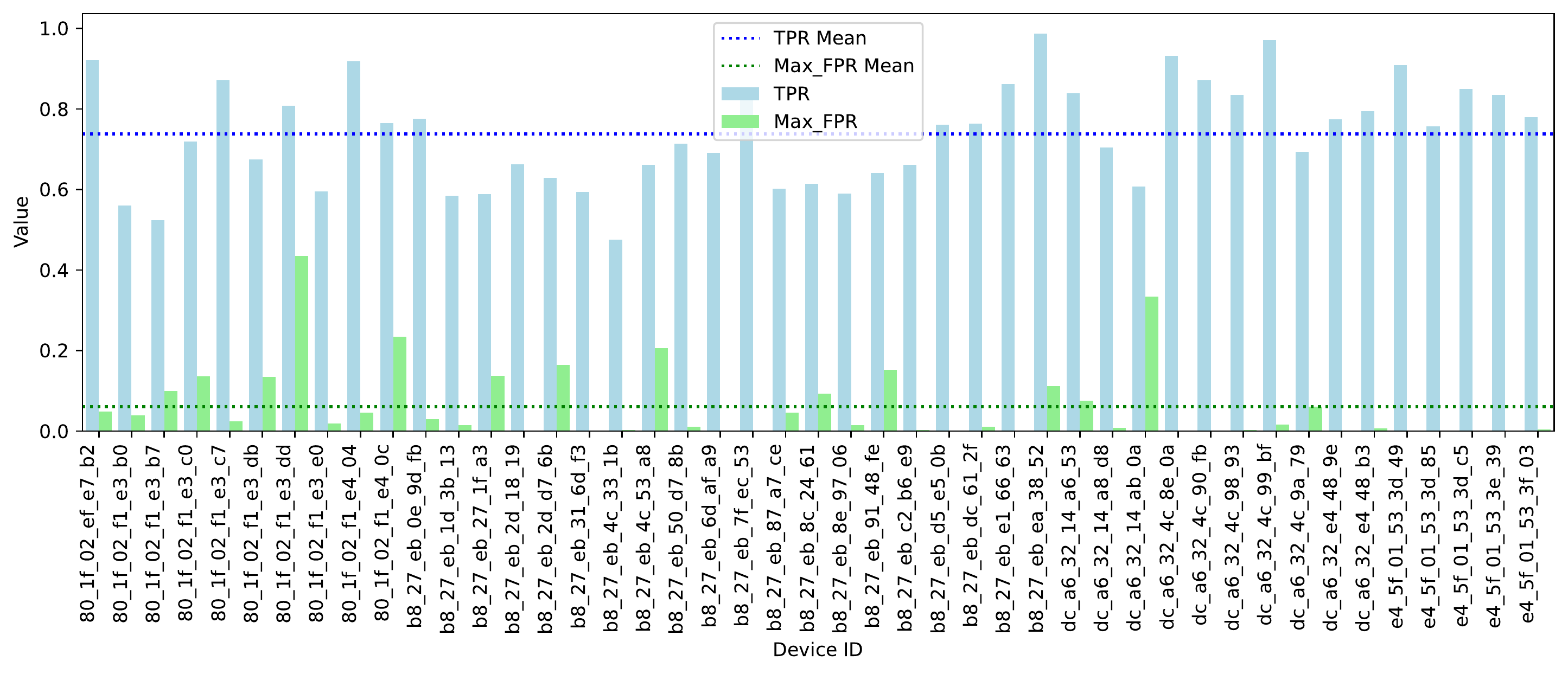}
    \caption{Transformer autoencoder TPR and maximum FPR comparison per device.}
    \label{fig:bars}
\end{figure*}

\figurename~\ref{fig:distributions} gives a closer look into the distributions of the TPRs and maximum FPRs of the 45 devices evaluated. It can be seen that both distributions are greatly separated, having only three cases where the maximum FPR goes over 0.20 and remains under 0.45. The TPR always stays over that value and reaches values close to 1 in some cases, having most of its values between 0.6 and 0.8. Besides, \figurename~\ref{fig:bars} shows the exact TPR and maximum FPR values for each one of the devices evaluated, having its MAC address as an identifier. In this graph can be observed that in the cases where the maximum FPR has a relatively high value (0.2 to 0.4), the TPR is way higher, guaranteeing that the authentication can be made reliably.

According to these results, a threshold-based authentication approach could be employed by the \textit{Authentication Module} to determine the result of the authentication process. An example can be a threshold for each device with a value 0.1 lower than the TPR achieved in the validation, as it is enough to differentiate all the devices present in the deployment.

The results achieved by the anomaly detection validation have demonstrated the feasibility of the proposed framework, as it was able to uniquely authenticate 45 single-board devices with identical hardware and software specifications. These findings point towards a promising direction for individual device authentication premised on hardware behavior, demonstrating the potential of Transformer models in this sphere.

\subsubsection{Resource Usage}

Although performance is the key characteristic to decide which model to use in the validation setup, resource usage during training and evaluation is also a critical point that should be taken into account when developing ML/DL-based solutions.

\tablename~\ref{tab:resources} shows the time and memory employed by the model. The training time statistics were collected using 10 epochs as the number of iterations over the training dataset. Besides, the evaluation time was obtained while evaluating the entire test dataset of the device. Finally, memory usage represents the size of the model after it has been completely trained.

\begin{table}[htpb!]
	\caption{Resource usage of each model (per device).}
	\centering
    \scriptsize
    \begin{tabular}{m{1.7cm}m{1.9cm}m{2cm}m{1.1cm}}
        \makecell[l]{\textbf{Model}} & \textbf{Training Time} &\textbf{Evaluation Time} & \textbf{Memory} \\
        \hline
        \hline
        1D-CNN & $\approx$47.79 s &  $\approx$1.44 s& 0.86 MB \\
        \hline
        LSTM & $\approx$283.68 s & $\approx$2.11 s & 1.33 MB \\
        \hline
        \makecell[l]{LSTM\_1D-CNN} & $\approx$306.92 s & $\approx$2.45 s & 1.83 MB \\
        \hline
        Transformer & $\approx$157.68 & $\approx$8.93 s & 7.77 MB \\
        \hline
    \end{tabular}
    \label{tab:resources}
\end{table}{}

Each model demonstrates distinct computational characteristics in terms of training time, evaluation time, and memory usage. The 1D-CNN model stands out as the most efficient, boasting the fastest training time of approximately 47.79 seconds and the quickest evaluation time of around 1.44 seconds. Additionally, it consumes the least amount of memory, using only about 0.86 MB. This combination of speed and efficiency makes it an appealing choice for resource-limited applications.

However, the LSTM model presents a significant increase in training time, taking approximately 283.68 seconds, and a slightly longer evaluation time of roughly 2.11 seconds. Coupled with a higher memory footprint of 1.33 MB, this model may demand greater computational resources than the 1D-CNN.

Interestingly, the hybrid LSTM+1D-CNN model exhibits the highest training time among the models, approximately 306.92 seconds, and has a considerable evaluation time of about 2.45 seconds. Its memory usage is also higher, at 1.83 MB, reflecting the complexity inherent to the combination of LSTM and 1D-CNN architectures.

Lastly, the Transformer model demonstrates a more moderate training time of approximately 157.68 seconds, albeit with the longest evaluation time of all models, around 8.93 seconds. More notably, it has a significantly higher memory usage, at a substantial 7.77 MB. While this may limit its applicability in memory-constrained environments, the Transformer model may excel in terms of capturing complex data patterns or delivering superior model accuracy, which are aspects not directly portrayed in the provided table.

In conclusion, while the 1D-CNN model is undeniably efficient regarding speed and memory usage, the Transformer models might offer better performance under certain circumstances. These trade-offs between time, memory usage, and potential model accuracy ought to be taken into account when deciding on the most suitable model for a particular scenario.

\addtxt{\subsection{Additional Validation in a simulated IoT deployment}}

\addtxt{Although the solution has already been validated in a real-world ElectroSense deployment, some additional challenges arise when adapting the framework to further scenarios. One of these challenges appears when new hardware models are present and hardware performance samples have to be collected from them. In this sense, gathering cycle counters from the device components is dependent on the exact hardware component, and the procedure might vary, requiring code adaptations in the data gathering process.}

\addtxt{ElectroSense is only compatible with Raspberry Pi hardware. Then, to explore this problem, an agriculture IoT scenario was simulated using nine additional devices from three new hardware models. Concretely, the list of devices employed was: 3 Rock64 devices, 3 RockPro64 devices, and 3 Orange Pi 2 Lite devices.}

\addtxt{The first step in the deployment process was to adapt the CPU and GPU cycle collection in the data gathering in order to be able to obtain the metrics described in \tablename~\ref{tab:features_dataset}. As the GPUs in these devices come from ARM Mali family, new counters should be leveraged. Concretely, the counter labeled as \textit{GPU\_ACTIVE} was chosen from the available options \cite{mali_counters}. The \textit{ARM HWCPipe} library was utilized for the collection of cycle counters \cite{HWCPipe}. In terms of CPU-based time collection, the \textit{perf} time was acquired similarly to the methodology for RPi devices, leveraging the \textit{perf\_counter\_ns()} function. The software to facilitate GPU task execution was modified from the source found in the \textit{ARM Compute Library} \cite{compute_library}.}

\addtxt{The data collection was executed for one week in these nine devices to have a large enough dataset for this secondary validation, around 4000 samples were collected per device in this period. Then, the data preprocessing steps were repeated as in the ElectroSense validation using Raspberry Pi devices, using a sliding window of 100 values and \textit{QuantileTransformer} normalization. After, the hyperparameter search for the Transformer models in charge of the authentication of each device was performed.}

\addtxt{\tablename~\ref{tab:second_validation} shows the authentication results from the nine devices employed in this validation. It can be seen that the results were even better than in the ElectroSense experiments. The TPR for all of the devices was above 0.90, while the maximum FPR stayed under 0.05 in all cases, enabling the threshold-based authentication as in the ElectroSense deployment. The improved results in this validation occurred due to the decreased number of devices from the same model being compared. In this sense, only three devices per model were present, so the similarities between the devices in the scenario were reduced.}

\begin{table}[htpb!]
    \centering
    \scriptsize
    {\addtable
    \caption{\addtxt{Validation in simulated IoT scenario}}
    \label{tab:second_validation}
    \begin{tabular}{cccc}
         \textbf{Device Model} & \textbf{Device (MAC)} & \textbf{TPR} & \textbf{Max. FPR} \\
         \hline
         \hline
         \multirow{3}{*}{RockPro64} & 86:a4:4c:5f:ff:95 & 0.8894 & 0.0417\\
         \cline{2-4}
         & 8a:32:38:8c:63:e6 & 0.9217 & 0.0131 \\
         \cline{2-4}
         & ee:db:54:9d:8a:67 & 0.9710 & 0.000 \\
         \hline
        \multirow{3}{*}{Rock64} & 42:58:a0:38:16:11 & 0.9094 & 0.007\\
         \cline{2-4}
         & 76:8f:be:c0:c5:3b & 0.9318 & 0.000 \\
         \cline{2-4}
         & 9a:1d:93:b3:b5:8f & 0.9187 & 0.041 \\
         \hline
        \multirow{3}{*}{Orange Pi 2 Lite} & c0:84:7d:82:4a:1e & 0.9653 & 0.011\\
         \cline{2-4}
         & c0:84:7d:82:1c:42 & 0.9142 & 0.000 \\
         \cline{2-4}
         & c0:84:7d:82:38:6d & 0.9760 & 0.002 \\
         \hline
    \end{tabular}
    }
\end{table}

\addtxt{This second validation approach serves as an example of how the proposed solution could be adapted to new scenarios where novel IoT device models are present, and code adaptions for data collection are necessary. It can be seen how the solution pipeline is still effective once the hardware monitoring is properly modified to gather the cycle counters from the monitored components. Besides, the proposal has been validated with some more devices, enhancing the scalability of the demonstrated solution.}

\section{DISCUSSION}
\label{sec:discussion}

This section outlines the limitations intrinsic to the suggested approach and provides essential understanding obtained from the research carried out. Through the series of tests performed in this work, coupled with a comparative analysis with existing literature, valuable observations and conclusions emerge. These findings serve not only as lessons gained but also highlight certain restrictions and limitations. The enumeration of lessons learned is as follows:

\textbf{Potential of Transformer Models for IoT Authentication.} The achieved results illustrate the innovative application of Transformer models in the field of hardware-based authentication. By employing a Transformer model, the framework was able to capture complex patterns in hardware behavior of each device, demonstrating a novel approach that could pave the way for future research in security and authentication. This lesson emphasizes the adaptability and potential of Transformer models in areas beyond natural language processing.

\textbf{Importance of Resource Usage Consideration}. The resource analysis emphasizes the critical consideration of resource usage during training and evaluation when developing Machine Learning or Deep Learning-based solutions. Different models demonstrated distinct computational characteristics in terms of training time, evaluation time, and memory usage. For example, the 1D-CNN model was found to be the most efficient, while the Transformer model had a significantly higher memory usage. These trade-offs between time, memory usage, and potential model accuracy must be carefully weighed when selecting the most suitable model for an IoT scenario where processing resources are limited.

\textbf{Versatility and Importance of Preprocessing Techniques} The paper emphasizes the importance of preprocessing techniques in handling time series data. The use of methods like grouping into vectors and data normalization (using QuantileTransformer) was essential in uncovering intricate trends within the data. This lesson serves as a reminder that preprocessing is not a one-size-fits-all step but a critical and adaptable component of the data analysis process, with significant implications for the success of the modeling and authentication framework.

\addtxt{\textbf{Adaptability to New Scenarios}. Based on the results of the second validation using additional IoT devices, it can be seen how only small changes in the data collection process are necessary to adapt the solution to the hardware of new devices. The remaining Transformer-based authentication pipeline remains functional in different scenarios and is hardware agnostic, enabling the application of the solution in a wide variety of environments as an additional security layer and complimenting traditional software-based authentication.}

Conversely, the subsequent constraints and limitations have been noted and warrant consideration in upcoming studies within this field:

\textbf{Determining Hardware Behavior Measurements and Hyperparameter Tuning.} The process of identifying the appropriate hardware behavior measurements or feature extraction for individual device authentication is complex and multifaceted. The implementation of the proposed methodology may necessitate multiple exploratory iterations to discover a combination that satisfies all the required properties in the generated fingerprint. Additionally, the Transformer model introduces further complexity due to the need for hyperparameter tuning. Finding the optimal set of hyperparameters for the Transformer model requires a meticulous search, adding to the trial-and-error nature of the process. This iterative analysis can be significantly minimized by examining the properties of the leveraged devices, including different components and operating frequencies, and by employing systematic hyperparameter optimization techniques. Since every chip inherently contains imperfections, the real challenge lies in devising accurate and effective methods to measure them and in fine-tuning the Transformer model to capture these unique characteristics. This complexity adds multiple layers of difficulty to the process and may require careful consideration, experimentation, and optimization to achieve the desired authentication accuracy.

\textbf{Training and Evaluation Time.} The varying training and evaluation times across different models, with the Transformer model exhibiting the longest evaluation time, present a limitation that may affect its suitability in time-sensitive applications. This constraint highlights the importance of considering both accuracy and computational efficiency in model selection and design.

\textbf{Threshold Setting for Anomaly Detection.} During validation, the anomaly detection threshold is set at 10\% of the reconstruction error in the training samples fed to the Transformer models. This choice of threshold might have specific implications on the sensitivity and specificity of the anomaly detection as it is manually assigned.
    
\addtxt{\textbf{Possible performance degradation over time.} As with any hardware, the components of IoT devices may undergo wear and tear, leading to gradual changes in their performance metrics. This natural aging process can alter the cycle skew and other performance parameters that the authentication system initially learned and recognized \cite{halak2016overview}. It has been experimentally verified that during the 100 days of data collection, the hardware performance has remained stable. To that end, the authentication experiments were repeated with different splits in the train/test data of each device, achieving very similar results no matter how the data was selected. However, longer periods might have a larger impact on hardware degradation.}

\section{CONCLUSIONS AND FUTURE WORK}
\label{sec:conclusions}

This paper proposes a framework for individual device authentication based on hardware behavior and outlier detection, which fundamentally relies on identifying inherent imperfections in the device chips. The framework, which leverages hardware behavior fingerprinting and Transformer autoencoders, establishes a unique 'fingerprint' for each device based on manufacturing imperfections in CPU, GPU, RAM, and Storage, even in those with identical specifications. These imperfections are modeled by generating a model trained with the "normal" data distribution of the hardware performance of each device. This provides a robust mechanism for device authentication, distinguishing between genuine and potentially harmful devices. The framework follows a modular design where device monitoring and security enforcement modules are deployed in the device and the data processing modules are hosted in a server with enhanced processing capabilities.

The practical implementation of this authentication framework in the ElectroSense platform demonstrates its effectiveness and real-world applicability. After 100 days of data collection using 45 Raspberry Pi devices, the Transformer-based autoencoder approach was implemented and compared with other state-of-the-art Deep Learning architectures such as LSTM and 1D-CNN for anomaly detection in time series. Despite the competitive performance of LSTM and 1D-CNN, the Transformer model emerged as the superior method, successfully authenticating all the devices. An average True Positive Rate (TPR) of 0.74$\pm$0.13 and an average maximum False Positive Rate (FPR) of 0.06$\pm$0.09 are achieved when performing one-versus-all authentication, a more complex task than the classification-based identification performed by other solutions in the literature. From these results, it can be concluded that the proposed approach not only prevents unauthorized device intrusions but also significantly contributes to the reliability of data analysis and the overall trustworthiness of the platform. 

Moving forward, this research line has room for future work and improvements. While the current study has focused on Raspberry Pi devices, further research should involve testing the proposed model with other IoT devices, expanding its scope, and ensuring its applicability across a broad range of hardware. In addition, the study has examined the model effectiveness primarily in the context of a spectrum crowdsensing platform, ElectroSense. Future investigations could explore its implementation in different types of crowdsensing applications, thereby contributing to a comprehensive understanding of the framework versatility.


\section*{Acknowledgment}
This work has been partially supported by \textit{(a)} the Swiss Federal Office for Defense Procurement (armasuisse) with the DEFENDIS and CyberForce (CYD-C-2020003)  projects and \textit{(b)} the University of Zürich UZH.

\bibliographystyle{abbrv}
\bibliography{references}

\end{document}